\documentclass[aps,prl, twocolumn, superscriptaddress, longbibliography]{revtex4-2}

\usepackage{times}
\usepackage{amsmath}
\usepackage{amssymb}
\usepackage{graphicx}
\usepackage[caption=false, position=top, singlelinecheck=off, justification=raggedright]{subfig}
\usepackage{color}
\usepackage{pst-node}
\usepackage[unicode]{hyperref}
\hypersetup{unicode=true, colorlinks=true, linkcolor=blue, citecolor=blue}

\begin{document}

\title{Observation of a phase transition from a continuous to a discrete time crystal}

\author{Phatthamon Kongkhambut}
\thanks{These two authors contributed equally}
\affiliation{Zentrum f\"ur Optische Quantentechnologien and Institut f\"ur Quantenphysik, Universit\"at Hamburg, 22761 Hamburg, Germany}

\author{Jayson G. Cosme}
\thanks{These two authors contributed equally}
\affiliation{National Institute of Physics, University of the Philippines, Diliman, Quezon City 1101, Philippines}

\author{Jim Skulte}
\affiliation{Zentrum f\"ur Optische Quantentechnologien and Institut f\"ur Quantenphysik, Universit\"at Hamburg, 22761 Hamburg, Germany}
\affiliation{The Hamburg Center for Ultrafast Imaging, 22761 Hamburg, Germany,}

\author{Michelle A. Moreno Armijos}
\affiliation{Instituto de F{\'i}­sica de S{\~a}o Carlos, Universidade de S{\~a}o Paulo, S{\~a}o Carlos, SP 13560-970, Brazil}

\author{Ludwig Mathey}
\affiliation{Zentrum f\"ur Optische Quantentechnologien and Institut f\"ur Quantenphysik, Universit\"at Hamburg, 22761 Hamburg, Germany}
\affiliation{The Hamburg Center for Ultrafast Imaging, 22761 Hamburg, Germany,}

\author{Andreas Hemmerich}
\thanks{hemmerich@physnet.uni-hamburg.de}
\affiliation{Zentrum f\"ur Optische Quantentechnologien and Institut f\"ur Quantenphysik, Universit\"at Hamburg, 22761 Hamburg, Germany}
\affiliation{The Hamburg Center for Ultrafast Imaging, 22761 Hamburg, Germany,}

\author{Hans Ke{\ss}ler}
\thanks{hkessler@physnet.uni-hamburg.de}
\affiliation{Zentrum f\"ur Optische Quantentechnologien and Institut f\"ur Quantenphysik, Universit\"at Hamburg, 22761 Hamburg, Germany}
\affiliation{Physikalisches Institut, Rheinische Friedrich-Wilhelms-Universit\"at , 53115  Bonn, Germany}

\date{\today}

\begin{abstract}   
Discrete (DTCs) and continuous time crystals (CTCs) are novel dynamical many-body states, that are characterized by robust self-sustained oscillations, emerging via spontaneous breaking of discrete or continuous time translation symmetry. DTCs are periodically driven systems that oscillate with a subharmonic of the external drive, while CTCs are continuously driven and oscillate with a frequency intrinsic to the system. Here, we explore a phase transition from a continuous time crystal to a discrete time crystal. A CTC with a characteristic oscillation frequency $\omega_\mathrm{CTC}$ is prepared in a continuously pumped atom-cavity system. Modulating the pump intensity of the CTC with a frequency $\omega_{\mathrm{dr}}$ close to $2\,\omega_\mathrm{CTC}$ leads to robust locking of $\omega_\mathrm{CTC}$ to $\omega_{\mathrm{dr}}/2$, and hence a DTC arises. This phase transition in a quantum many-body system is related to subharmonic injection locking of non-linear mechanical and electronic oscillators or lasers.
\end{abstract}

\maketitle 

\section{Introduction}
The conceptual idea of time crystals (TCs) was first described as a self-sustaining oscillatory behavior in biological systems Ref.~\cite{Win:80}, and then established as a dynamical many-body state in physical systems in Refs.~\cite{Wilczek2012, Shapere2017}. A defining feature of these states is the spontaneous breaking of discrete or continuous time translation symmetry, giving rise to robust oscillatory motion in an extended region of their parameter space. Two distinct  scenarios for the emergence of these states are as follows: Firstly, for closed systems, the continuous time translation symmetry (CTTS) can be explicitly broken by a periodic external drive and the remaining discrete time translation symmetry (DTTS) is spontaneously broken by an oscillatory response of the system with a period longer than that of the drive. An ergodicity slowdown mechanism prevents the system from heating up to an infinite temperature for long times \cite{mbl, Zal:23}. This scenario is referred to as a "discrete TC" (DTC). Secondly, a TC state can also arise for open systems coupled to a bath. Similar to the DTC in closed systems, a periodic drive triggers a subharmonic oscillatory motion of the system, resulting in a dissipative DTC. The appropriately designed bath can act as a sink for the entropy produced by the system. \cite{rao:2016}. We note that, in contrast to closed systems, in open systems, a TC can also emerge in the absence of periodic driving, resulting in the spontaneous breaking of CTTS. This dynamical state is referred to as a ``continuous TC'' (CTC) \cite{Iemini2018, Kessler2019, Buc:19}.

\begin{figure*}[!htpb]
\centering
\includegraphics[width=2.0\columnwidth]{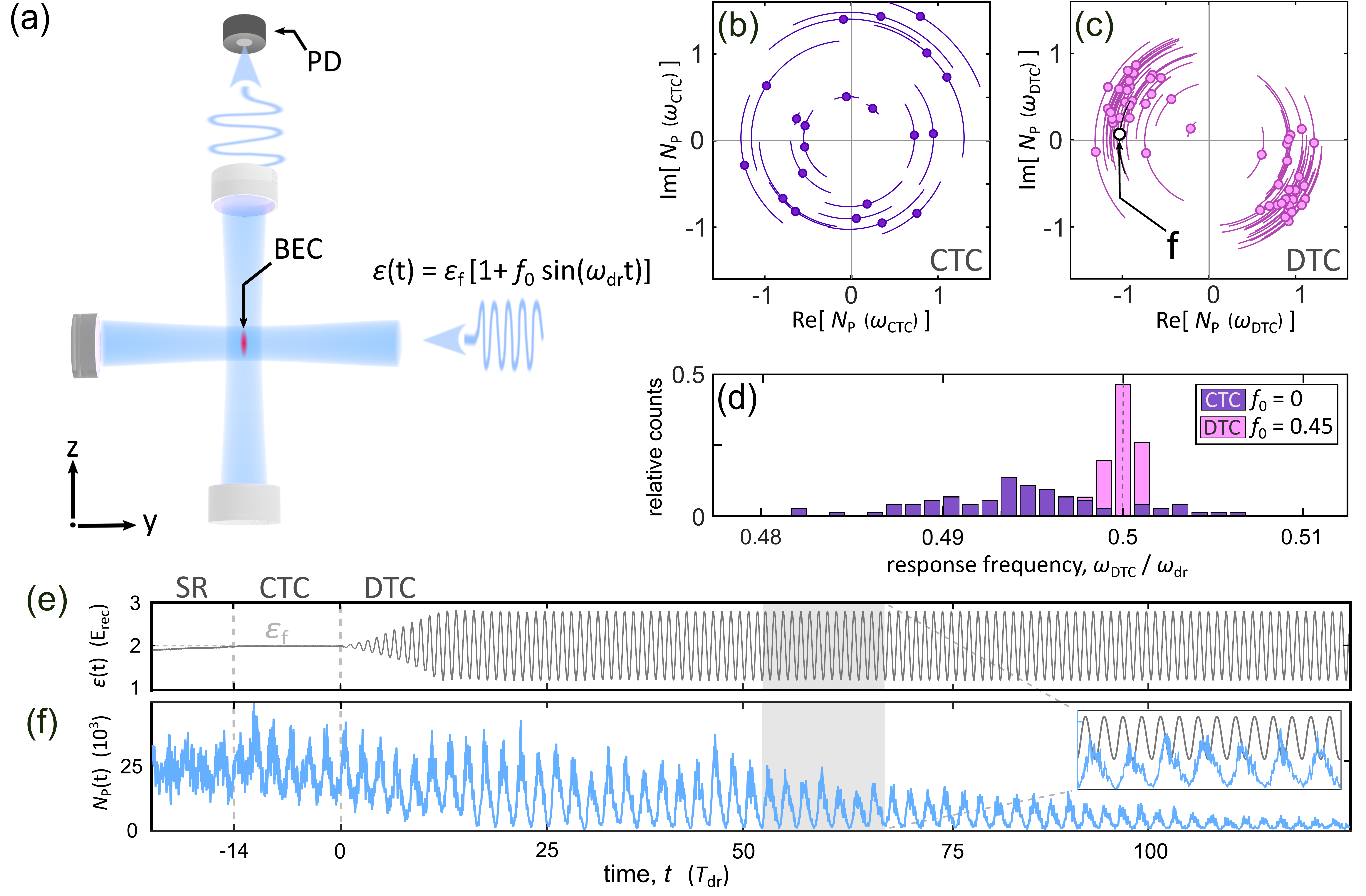}
\caption{(a) Schematic sketch of the atom-cavity system psystem, which is periodically pumped by an optical standing wave potential transverse to the cavity axis. (b) and (c) Distributions of the time phase intra-cavity photon number $N_\mathrm{P}(t)$ oscillating with the main response frequencies $\omega_\mathrm{CTC}$ in the CTC case shown in (b) and $\omega_\mathrm{DTC}$ in the DTC case shown in (c). The error bars represent the phase uncertainty within the discrete FFT resolution of $100$ Hz. The uncertainty with regard to the radial dimension, i.e. the amplitude uncertainty, is negligibly small. Note that the tilt of the two observed phase values by an angle of about $\pi/4$ with respect to the modulation signal is due to the retardation of the cavity field dynamics, which is caused by the small cavity bandwidth.. (d) Histogram of the relative number of counts of the response frequency $\omega_\mathrm{DTC}$ in units of the driving frequency $\omega_\mathrm{dr} = 2\pi\times 22.5\,$kHz for the non-modulated ($f_0=0$) case in purple (dark) and the modulated case ($f_0=0.45$) in pink (bright), respectively. Here, the same data is used as for (b) and (c), respectively. (e) Pump protocol and (f) evolution of $N_\mathrm{P}(t)$ for a typical experimental realization. Below $t=-14\,T_\mathrm{dr}$ (first dashed vertical line), $\varepsilon(t)$ is ramped up while the systems is in the superradiant (SR) phase, indicated by a non-zero, non-oscillatory $N_\mathrm{P}(t)$. Between $t=-14\,T_\mathrm{dr}$ and $t=0\,T_\mathrm{dr}$, $\varepsilon(t)$ reaches a critical value $\varepsilon_{\mathrm{f}}$ and the CTC phase arises, displayed by an oscillatory $N_\mathrm{P}(t)$. Above $t=0\,T_\mathrm{dr}$ (second dashed vertical line), modulation results in a DTC, indicated by an oscillatory $N_\mathrm{P}(t)$ with a significantly lower bandwidth than that of the CTC (cf. d). The inset in (f) shows a zoom of $\varepsilon(t)$ and $N_\mathrm{P}(t)$ for the time interval marked by the gray rectangle. The effective cavity pump detuning is $\delta_\mathrm{eff}=-2\pi~\times8.2$~kHz and the final pump strength $\varepsilon_\mathrm{f}=2.0~E_\mathrm{rec}$ for all measurements presented in the main text.}
\label{fig:1} 
\end{figure*}

The theoretical conceptualization of TCs in the context of many-body physics was followed by rapid experimental progress. DTCs in nearly closed systems have been realized in arrays of trapped ions, nitrogen vacancy centers, and in a mechanically kicked Bose-Einstein condensate (BEC) \cite{Zhang2017, Choi2017, smits18}. Discrete dissipative TCs were demonstrated in a BEC of neutral atoms in an optical cavity \cite{Kessler2021, Kon:21, zhu19, Skulte2021} and in an optical microcavity filled with a Kerr medium \cite{Tah:22}. Finally, continuous dissipative TCs were, for example, realized in magnon BECs \cite{Aut:18}, BECs of neutral atoms \cite{Kon:22}, in collections of spins in a semi-conductor matrix \cite{Greilich2024}, in photonic metamaterials \cite{Liu:23}, or doped crystals \cite{Chen2023}.

\begin{figure}[!htpb]
\centering
\includegraphics[width=1\columnwidth]{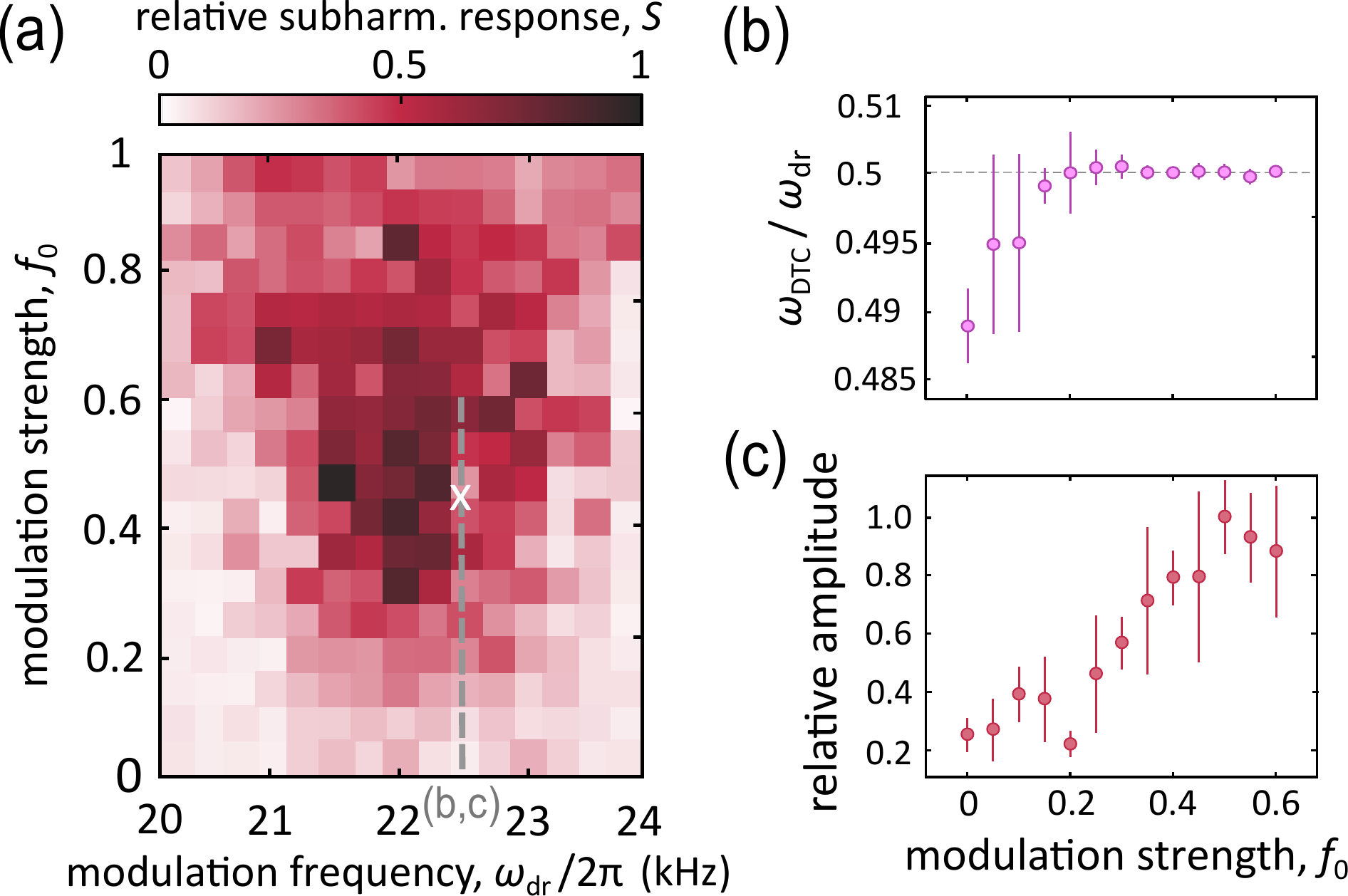}
\caption{(a) Relative subharmonic response $S$ versus driving strength $f_0$ and frequency $\omega_\mathrm{dr}$ for fixed effective detuning $\delta_\mathrm{eff}=-2\pi~\times8.2$~kHz and final pump strength $\varepsilon_\mathrm{f}=2.0~E_\mathrm{rec}$. To obtain (a), we ramped the pump strength $\varepsilon(t)$ to its final value $\varepsilon_\mathrm{f}$ for fixed $\delta_\mathrm{eff}$. After a $0.5$~ms long hold time, the driving strength is ramped to its desired value $f_\mathrm{0}$ for a selected driving frequency $\omega_\mathrm{dr}$ within $0.5$~ms and subsequently held constant for 10~ms. The parameter space is divided into 15 $\times$ 18 plaquettes and averaged over 5 to 10 experimental realizations. The white cross indicates the parameter values $f_\mathrm{0}=0.45\,$kHz and $\omega_\mathrm{dr}= 2\pi\times22.5\,$kHz, which are used for the measurements in Figs.~\ref{fig:1}(c-f), Fig.~\ref{fig:3} and Fig.~\ref{fig:4}. (b) Response frequency $\omega_\mathrm{DTC}$ in units of the driving frequency $\omega_\mathrm{dr}$, plotted versus the driving strength $f_0$. $\omega_\mathrm{DTC}$ is extracted as the position of a Gaussian fit to the Fourier spectrum of the intra-cavity photon number $N_\mathrm{P}(t)$. (c) Relative amplitude of the main spectral component at frequency $\omega_\mathrm{DTC}$, plotted versus $f_0$. The plots in (b) and (c) correspond to the path marked in (a) by the gray dashed line.  The error bars show the standard deviation and hence represent the shot-to-shot fluctuations.}
\label{fig:2} 
\end{figure}

Injection locking (IL) is a phenomenon, which can arise if a nonlinear dissipative oscillator in a limit-cycle state \cite{Str:18}, that is driven externally with a driving frequency $\omega_\mathrm{dr}$. For sufficiently strong driving, the oscillator locks to the external drive. This locking can occur at the driving frequency itself, or, more generally, at a rational ratio of the driving frequency \cite{Jen:83, Jen:84}. A specific case is subharmonic IL, in which the phase-locking occurs at an integer fraction of the driving frequency, i.e. $\omega_\mathrm{dr} / n$ with $n \in \{ 1,2,...\}$. We note that IL is a key phenomenon in electronic circuits, laser systems, and biological systems, such as the circadian rhythms of organisms \cite{Daan2001} or the synchronization of flashing of fireflies exposed to a periodically switching torch \cite{Hanson1971}. In biological systems or mathematical science this phenomenon is referred to as entrainment \cite{Str:18}.

\section{Results}
In this article, we demonstrate subharmonic IL in the context of time crystals. Here, a limit cycle is provided by a CTC produced in an atom-cavity system, oscillating at a frequency $\omega_\mathrm{CTC}$. We drive the system with a perturbation with a frequency $\omega_\mathrm{dr}$, which is close to $2 \,\omega_\mathrm{CTC}$. As a result, the CTC locks to the driving frequency, performing an oscillatory motion at $\omega_\mathrm{dr} / 2$, i.e. at a subharmonic frequency. In the language of time crystals, we realize a non-equilibrium phase transition between a CTC and DTC. In the terminology of laser physics, we establish subharmonic IL in a quantum many-body system.

Our setup is shown in Fig.~\ref{fig:1}(a). We start with a CTC prepared in an atom-cavity system (cf. Fig.~\ref{fig:1}(a)) consisting of a BEC located in a high-finesse optical cavity, pumped transversally by an optical standing wave at constant intensity. As reported in Ref.~\cite{Kon:22}, this leads to robust self-sustained oscillations of the intra-cavity photon number $N_\mathrm{P}(t)$, which establishes a CTC. Its frequency $\omega_\mathrm{CTC}$ can be associated with the emergence of a limit cycle \cite{Pia:15, Kessler2019, Kessler2020, Colella2022, Nie2023, Skulte2024}. As seen in Fig.~\ref{fig:1}(b), the oscillation of the CTC breaks the CTTS. The real and imaginary parts of the Fourier spectrum of $N_\mathrm{P}(t)$ at the dominant frequency $\omega_\mathrm{CTC}$ are plotted here for different experimental implementations. The phase values of the Fourier spectra are randomly distributed between $0$ and $2\pi$, confirming the expected spontaneous breaking of CTTS. 

Next, we modulate the intensity of the pump field $\varepsilon(t)$ at a frequency $\omega_\mathrm{dr}$ close to $2\,\omega_\mathrm{CTC}$. The periodic drive breaks the CTTS of the atom-cavity platform such that the modulated system only retains DTTS. Under the influence of the modulation, the system converts into a DTC (cf. Ref.~\cite{Kon:21}) with an oscillation frequency $\omega_\mathrm{DTC}$ approaching $\omega_\mathrm{dr}/2$ for sufficiently strong driving. In Fig.~\ref{fig:1}(c), we analyze the Fourier spectra at the emission frequency $\omega_\mathrm{DTC}$ for different experimental implementations. Only two almost equiprobable (49\% and 51\%) phase values, approximately differing by $\pi$, are observed, confirming spontaneous breaking of the DTTS (cf. video in the appendix). The modulation, in addition to the observed frequency pulling towards subharmonic response, also gives rise to a line narrowing of the DTC emission as presented in the histogram in Fig.~\ref{fig:1}(d). This is also seen in Fig.~\ref{fig:1}(e) and (f), showing that the oscillations observed in $N_\mathrm{P}(t)$ become more regular as the modulation sets in at $t=0$ and the lifetime of the TC extends to more than a hundred driving cycles. The life time in both regimes, CTC and DTC, is mainly limited by atom loss from the trap and the associated decrease of the collective atom-light coupling.

We note that the transition of the CTC to the DTC occurs in two steps, as $f_0$ is increased from zero to a value above the critical value of the DTC phase. As we demonstrate in the Appendix (cf. Fig.~\ref{sfig:11}), the intermediate regime is a quasicrystalline state \cite{Aut:18,Giergiel2018}, in which the limit cycle dynamics of the CTC state transitions to a limit torus dynamics. In the frequency representation $N_\mathrm{P}(\omega)$, this transitions manifests itself as side-bands close to the dominant frequency peak $\omega_{\mathrm{CTC}}$ at frequencies $\omega_{\mathrm{CTC}} \pm (\omega_{\mathrm{CTC}} - \omega_{\mathrm{dr}} / 2$. When $f_0$ is further increased above a critical value of the DTC phase, the dominant emission is shifted towards $\omega_{\mathrm{dr}} / 2)$ and all side bands disappear. Hence, a phase transition towards a DTC arises. We find this intermediate limit torus regime not to be detectable experimentally, in the regime and in the experimental setup used here, due to the finite lifetime of the atoms, and other imperfections that limit the frequency resolution.

\section{Methods}
The experimental set-up consists of a Bose-Einstein condensate (BEC) of $N_\mathrm{a}=4\times10^4$ $^{87}$Rb atoms strongly coupled to a single mode of an optical high-finesse cavity. The system is pumped transversally, perpendicular to the cavity axis at a wavelength of $\lambda_\mathrm{p} = 791.59\,$nm (cf. Fig.~\ref{fig:1}(a)). The pump light is blue detuned with respect to the relevant atomic transition, the $D_1$-line of $^{87}$Rb at $794.98\,$nm. The effective pump-cavity detuning is chosen to be negative for all experiments presented here and is defined as $\delta_{\mathrm{eff}}\equiv \delta_{c}-\delta_-$, where $\delta_c \equiv \omega_\mathrm{p} - \omega_\mathrm{c}$ is the detuning between the pump field frequency $\omega_\mathrm{p}$ and the cavity resonance frequency $\omega_\mathrm{c}$, and $\delta_- = \frac{1}{2}N_\mathrm{a} U_0$ denotes the collective light shift of the cavity resonance caused by the atomic ensemble for the relevant left circular polarisation mode of the cavity. For the chosen pump wavelength $\lambda_\mathrm{p}$, the light shift per photon is $U_\mathrm{0} = 2\pi\times0.7\,$Hz. The cavity operates in the recoil resolved regime, meaning that the field decay rate of the cavity $\kappa=2\pi\times 3.2$~kHz, which sets the time scale for the intra-cavity light field dynamics, is comparable to the recoil frequency $\omega_\mathrm{rec}=2\pi \times 3.7$~kHz. The latter sets the time scale for the density distribution of the BEC to adapt to changes of the intra-cavity light field \cite{Kessler2014, Klinder2016}. This unique regime is a key prerequisite for the existence of the time crystalline phases \cite{Kessler2021, Kon:21, Kon:22}, which are the starting point of the work presented here. The experimental cycle starts with preparing a CTC. For this, we first prepare the superradiant (SR) phase \cite{Black2003, Baumann2010, Klinder2015, Klinder2016} by linearly increasing the pump-field strength $\varepsilon(t)$. When $\varepsilon$ exceeds a critical value, we observe a non-zero intra-cavity photon number $N_\mathrm{P}$, indicating the formation of the SR phase, in which the atoms self-organize to form a density wave that enables superradiant scattering of pump light into the cavity mode. The phase transition to the SR phase goes along with spontaneous breaking of a $\mathbb{Z}_2$ translation symmetry in space \cite{baumann2011}. Increasing $\varepsilon(t)$ further and holding it at a constant value $\varepsilon = \varepsilon_\mathrm{f}$, for appropriate settings of $\delta_\mathrm{eff}$ and $\varepsilon_\mathrm{f}$, causes the system to develop periodic motion, corresponding to a CTC \cite{Kon:22}. Subsequently, the pump strength is modulated according to $\varepsilon(t)=\varepsilon_\mathrm{f} \, [1 + f_0\cos(\omega_\mathrm{dr}t)]$, with the mean pump strength $\varepsilon_\mathrm{f}$, driving strength $f_0$, and frequency $\omega_\mathrm{dr}$. If the driving strength $f_0$ is sufficiently large, the response frequency $\omega_\mathrm{DTC}$ locks to the first subharmonic of $\omega_\mathrm{dr}$ and a DTC is realized (cf. Fig.~\ref{fig:1}(e,f)). Note that, in contrast to the DTC observed in reference \cite{Kessler2021}, here the system does not periodically switch back and forth between the two density gratings associated with the two symmetry broken states of the SR phase but spontaneously chooses one or the other.
\begin{figure}[!tpb]
\centering
\includegraphics[width=1\columnwidth]{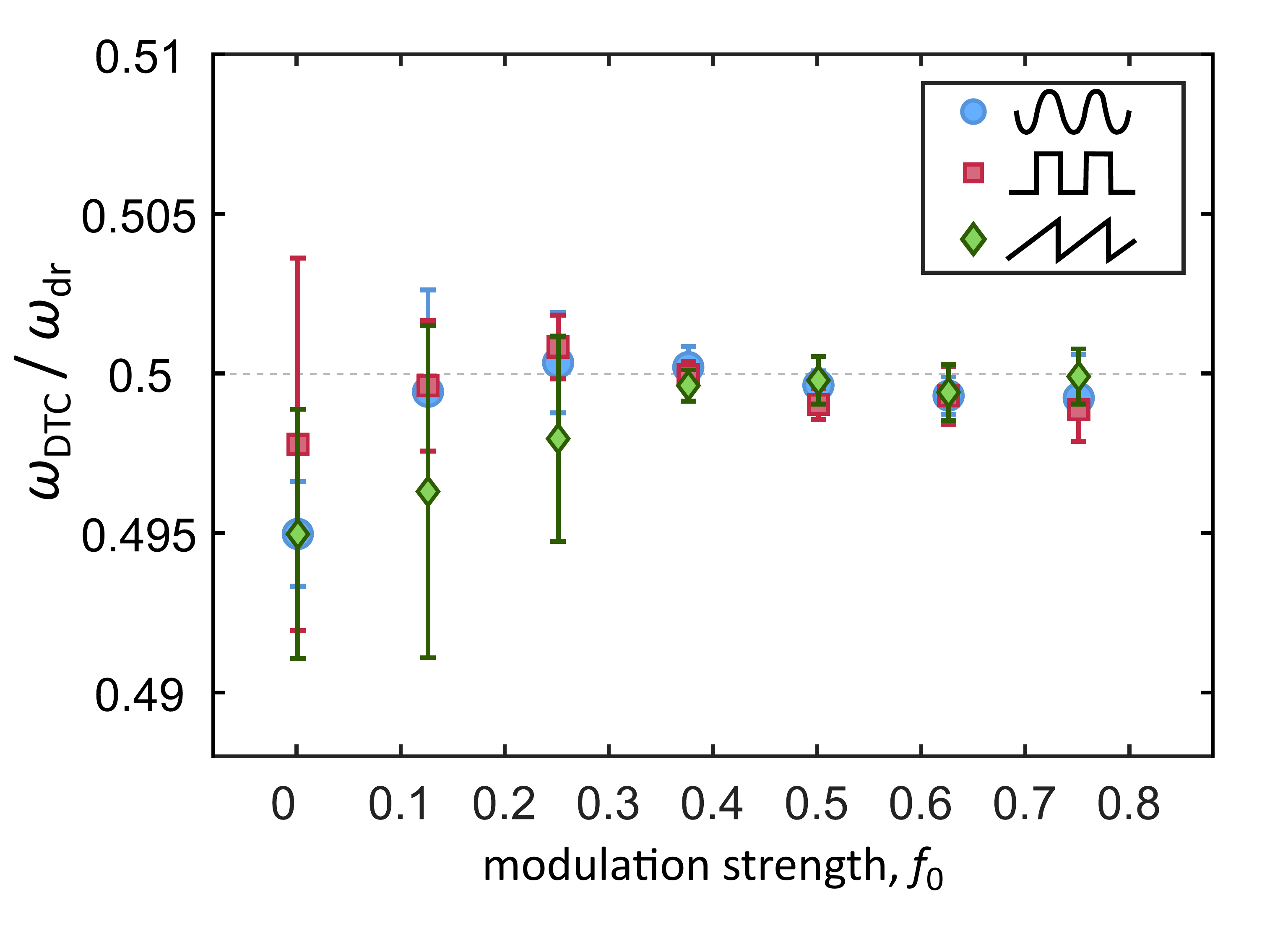}
\caption{The blue markers show the response frequency $\omega_\mathrm{DTC}$ in units of the driving frequency $\omega_\mathrm{dr}$ plotted versus the driving strength $f_0$ for a sinusoidal modulation waveform as used for the measurements presented in Figs.~\ref{fig:1},\ref{fig:2}, and \ref{fig:4}. The error bars show the standard deviation and hence represent the shot-to-shot fluctuations, which are strongly suppressed with increasing values of $f_{0}$. The red (square) and green (diamond) markers show the cases of modulation with square wave or sawtooth waveforms, respectively. The experimental protocol is the same as for the measurements in Figs.~\ref{fig:2}(b,c).}
\label{fig:3} 
\end{figure}

\begin{figure*}[!htpb]
\centering
\includegraphics[width=2\columnwidth]{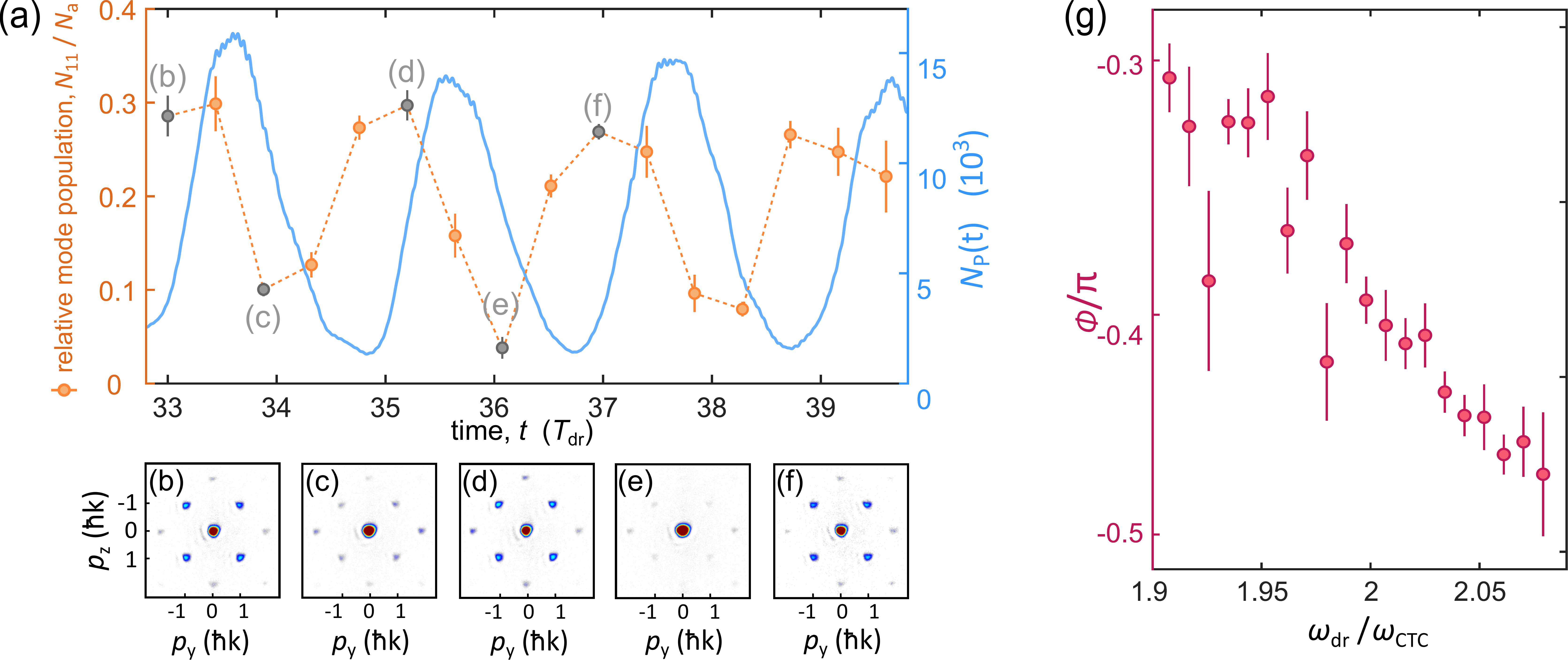}
\caption{(a) Blue solid line: intra-cavity photon number $N_\mathrm{P}(t)$ averaged over five experimental realizations. Orange and gray markers: $N_\mathrm{11} / N_\mathrm{a}$, sum of the populations of the four momentum modes $\{p_\mathrm{y},p_\mathrm{z}\} = \{\pm 1,\pm 1\} \hbar k$ normalized to the total atom number $N_\mathrm{a}$. The dashed orange line connects the data points to guide the eyes. (b-f) Averaged momentum spectra used to obtain the data points marked in (a) by the symbols highlighted in black. We chose the same parameters as for the data presented in Fig.~\ref{fig:1}. Since the system spontaneously picks one of the two possible phases of the DTC state, we first extract the time phase as in Fig.~\ref{fig:1}(c) from a Fourier spectrum and then post-select realizations with similar phase values before averaging. More details about the post-selection process are found in the appendix together with a video showing the time evolution of the momentum spectra. (g) Phases of the oscillations of $N_\mathrm{P}(t)$. The error bars show statistical errors for averaging over multiple realizations.}
\label{fig:4} 
\end{figure*}

\section{Discussion}
As a first experiment, we identify the optimal parameter values of $f_\mathrm{0}$ and $\omega_\mathrm{dr}$ where the IL of the CTC works most efficiently. We fix the effective detuning $\delta_\mathrm{eff} = -2\pi\times8.2\,$kHz and the final pump strength $\varepsilon_\mathrm{f} = 2.0\,E_\mathrm{rec}$. For these parameters, we observed the strongest subharmonic response while keeping $f_\mathrm{0}$ and $\omega_\mathrm{dr}$ fixed (cf. Fig.~\ref{sfig:1} in the appendix). The protocol used for the measurement presented in Fig.~\ref{fig:2} is as follows: We linearly increase the pump strength $\varepsilon(t)$ to its desired final value $\varepsilon_\mathrm{f}=2.0\,E_\mathrm{rec}$ for fixed $\delta_{\mathrm{eff}}=-2\pi\times8.2\,$kHz. This is followed by a waiting time and a linear ramp-up of the driving strength $f_\mathrm{0}$, both with a duration of $0.5\,$ms. Then, we hold all the pump parameters constant, record $N_\mathrm{P}(t)$ during 10~ms, and calculate the Fourier transform $N_\mathrm{P}(\omega)$ of $N_\mathrm{P}(t)$ (using a Fast Fourier transform method (FFT)). To quantify the degree of IL, we extract the subharmonic response $S = N_\mathrm{P}(\omega_\mathrm{dr} / 2) / \mathrm{Max}_{\{\omega_\mathrm{dr}, f_0\} }  [ N_\mathrm{P}(\omega_\mathrm{dr} / 2)]$, which is the amplitude of the single-sided spectrum at half of the driving frequency $N_\mathrm{P}(\omega_\mathrm{dr} / 2)$, normalized to its maximal value observed across the considered portion of the $\{ \omega_\mathrm{dr}, f_\mathrm{0} \}$-space.

In Fig.~\ref{fig:2}(a), we observe a large area showing a strong subharmonic response $S$. For the optimal choice of $\delta_\mathrm{eff}$ and $\varepsilon_\mathrm{f}$ (see appendix), $S$ is increased more than fourfold when compared to its value without modulation. The maximal value of $S$ arises for a driving frequency $\omega_\mathrm{dr}$ close to twice the CTC frequency $\omega_\mathrm{CTC}$, where $\omega_\mathrm{CTC} \approx 2 \pi \times11\,$kHz for the optimal choice of $\delta_\mathrm{eff}$ and $\varepsilon_\mathrm{f}$. The optimal driving strength of about $f_\mathrm{0}=0.45$ exceeds the value predicted in our simulations (cf. appendix \textcolor[rgb]{0.58,0,0}{Fig.\ref{sfig:6}(a)}), which may be attributed to the limited experimental lifetime and the contact interaction of the BEC, which is not accounted for in the calculations. For increasing $f_0$, the synchronization happens faster and is more robust in the sense that larger values of the subharmonic response $S$ are observed together with an extension over longer time periods. Based upon the observation of spontaneous breaking of DTTS (cf. Fig.~\ref{fig:1}(d)) and robustness of the subharmonic response against temporal perturbations of all four pump and modulation parameters ($\delta_{\mathrm{eff}}, \varepsilon_\mathrm{f}, \omega_\mathrm{dr}, f_0$), we claim to observe a transition between a CTC and a DTC (see appendix for details). We investigate this transition further for a fixed driving frequency $\omega_\mathrm{dr}=2\pi\times22.5\,$kHz. For each experimental realization, we obtain the Fourier spectrum as described above, but instead of considering its amplitude at $\omega_\mathrm{dr}/2$, we fit a Gaussian to extract the dominant response frequency $\omega_\mathrm{DTC}$ as the frequency at the maximum of the Gaussian and its corresponding amplitude. These quantities are plotted versus the driving strength $f_\mathrm{0}$ in Figs.~\ref{fig:2}(b) and (c), respectively. For increasing $f_0$, the response frequency $\omega_\mathrm{DTC}$ approaches the value $\omega_\mathrm{dr}/2$. Each data point is an average of around ten experimental realizations and the error bars in Figs.~\ref{fig:2}(b,c) indicate the standard deviation, representing shot-to-shot fluctuations. These fluctuations are due to atom number variations in the BEC, originating from a combination of inherent quantum noise and technical instabilities. Interestingly, we find that for sufficiently strong driving, the emergence of the DTC is accompanied by a strong suppression of the shot-to-shot fluctuations of $\omega_\mathrm{DTC}$ (cf. Fig.~\ref{fig:2}(b)), while at the same time, the relative amplitude of the dominant spectral component at frequency $\omega_\mathrm{DTC}$ increases by almost a factor of 5 (cf. Fig.~\ref{fig:2}(c)).

To further assess the efficiency of the IL process with respect to frequency pulling and locking, we plot in Fig.~\ref{fig:3} the response frequency $\omega_\mathrm{DTC}$, averaged over about ten experimental realizations, against the driving strength $f_0$, using three different modulation waveforms: sinusoidal (blue circles), square wave (red squares), and sawtooth (green diamonds). The protocol is otherwise the same as the one described in the previous paragraph. For all three waveforms, $\omega_\mathrm{DTC}$ is pulled towards $\omega_\mathrm{dr}/2$ for sufficiently strong driving strength and we observe a plateauing of $\omega_\mathrm{DTC}$ above $f_\mathrm{0}\approx 0.3$. The shot-to-shot fluctuations, given by the error bars, are seen to significantly decrease as the response locks onto the subharmonic of the drive. Moreover, frequency locking is reached for smaller $f_\mathrm{0}$ when using a square wave or sinusoidal modulation when compared to a sawtooth modulation. This may be explained as follows: the modulation is implemented as $\varepsilon(t)=\varepsilon_\mathrm{f} \, [1 + f_0 \, g(t)]$, where $g(t)$ denotes one of the three waveforms oscillating between the maximal and minimal values $1$ and $-1$. For this specification of $f_0$, the amplitude of the fundamental harmonic contribution of the square, sinusoidal, and sawtooth waveforms are $\{4/\pi,1,2/\pi \}$, respectively. Hence, when compared to the sinusoidal waveform, the square wave and the sawtooth modulation should produce tighter or weaker locking, respectively. For the square wave, however, the higher harmonic components give rise to increased heating, which reduces the atom-cavity coupling and hence acts to compensate for the tighter locking.

So far, we have restricted ourselves to obtaining information about the atom-cavity system by analyzing the light field leaking out of the cavity, which serves as a non-destructive monitor for the light-matter dynamics. However, we also have direct access to the matter sector via momentum spectra measured after a $25\,$ms long free expansion of the ensemble. This  time-of-flight (TOF) technique is destructive, and we need to prepare a new matter sample every time a momentum spectrum is recorded. In the CTC phase, in each experimental realization, the intra-cavity light field and the corresponding matter grating oscillate with a random time phase as a consequence of CTTS breaking (cf. Fig.~\ref{fig:1}(b)). Hence, averaging over multiple realizations, in order to improve signal-to-noise, washes out the dynamical signatures of the observed momentum distributions. In the DTC regime, only two time phases, differing by $\pi$, emerge. These phases can be discriminated by analyzing Fourier spectra according to Fig.~\ref{fig:1}(c), such that post-selection allows for averaging momentum spectra with the same phase value. With this, we directly observe the dynamics of the atomic matter grating. In Fig.~\ref{fig:4}(a), the time evolution of $N_\mathrm{P}(t)$ is plotted as a solid blue line, and the time evolution of the sum $N_\mathrm{11}$ of the populations of the four momentum modes $\{p_\mathrm{y},p_\mathrm{z}\} = \{\pm 1,\pm 1\}\hbar k$, normalized to the total atom number $N_\mathrm{a}$, is shown by orange and gray markers. In order to obtain $N_\mathrm{11} / N_\mathrm{a}$, momentum spectra like those shown in Figs.~\ref{fig:4}(b-f) are recorded, post-selected to only account for similar time phase values, and averaged. We observe an oscillation in the dynamics of $N_\mathrm{11} / N_\mathrm{a}$ at a frequency similar to that of the intra-cavity photon number but notably with a time phase shifted relative to the time phase of $N_\mathrm{P}(t)$. This retardation between the dynamics of the light field and the matter distribution is a key feature of our recoil resolved atom-cavity system \cite{Kessler2016, Georges2017} and is consistent with simulations using an idealized model for the atom-cavity system (cf. appendix). In Fig.~\ref{fig:4}(g), the phase of the oscillation of $N_\mathrm{P}(t)$ with respect to the phase of the drive is plotted versus $\omega_{\mathrm{dr}}$, which is tuned across the resonance $\omega_{\mathrm{dr}} = 2\,\omega_{\mathrm{CTC}}$. The observed dissipation-induced change of the phase, when $\omega_{\mathrm{dr}}$ is varied, is a characteristic signature of IL or entrainment. The nearly linear decrease with a negative slope is reproduced by the simulations in the appendix. 
\\

\section{Conclusion}
In conclusion, we have demonstrated dynamical control of a phase transition between two time crystalline phases. Taking the continuous time crystalline phase of a transversally pumped atom-cavity system as a starting point, we have applied external driving at a frequency of approximately twice the frequency of the continuous time crystal. For sufficiently strong driving, the system locks to the external drive in a subharmonic manner, resulting in a discrete time crystal. This phenomenon establishes subharmonic IL of limit cycles of a nonlinear dissipative oscillator in the context of many-body systems. Therefore, we establish a non-trivial interface between classical non-linear dynamics and time crystals, which suggests a vast range of dynamical phenomena to be understood and established in time crystals and related dynamical many-body states \cite{Bang2024,He2024}. 

\section{Data availability statement}
All data presented in this article can be provided by the authors A.H. and H.K. upon request.

\section{Acknowledgments}
\begin{acknowledgments}
We thank J. Klinder, C. Ni, A. B{\"o}lian, and C. Georges for their support during the early stage of the project. A.H. acknowledges useful discussions with C. Zimmermann and J. Marino. J.G.C  thanks R.J.L. Tuquero for helpful discussions. P.K. thanks M. Sauer for assembling video contents in the supplementary material. A.H. acknowledges support by the QuantERA II Programme that has received funding from the European Union's Horizon 2020 research and innovation programme under Grant Agreement No.~101017733. P.K., J.S., L.M. and A.H. acknowledge the Deutsche Forschungsgemeinschaft (DFG) for funding through SFB-925 - project~170620586, and the Cluster of Excellence "`Advanced Imaging of Matter"' (EXC 2056) - project No.~390715994. J.G.C. acknowledges funding from the UP System Balik PhD Program (OVPAA-BPhD-2021-04). J.S. acknowledges support from the German Academic Scholarship Foundation. H.K. acknowledges funding by the state of North Rhine-Westphalia through the EIN Quantum NRW program and by the Deutsche Forschungsgemeinschaft (DFG) through grant DFG-KE 2481/1-1. L.M. acknowledges  co-funding by ERDF of the European Union and by ``Fonds of the Hamburg Ministry of Science, Research, Equalities and Districts (BWFGB)''.
\end{acknowledgments}

Note: During peer-review of this manuscript, we got to know about related experiments using a coupled electron-nuclear spin system  \cite{Greilich2024_2}.

\section{References}
\bibliography{references_entrainment}

\appendix

\section*{Appendix}
\setcounter{section}{1}

\section{Experimental details}
The experimental setup, as sketched in Fig.~1(a) in the main text, is comprised of a magnetically trapped BEC of $N_\mathrm{a} = 4\times 10^4$  $^{87}$Rb atoms, dispersively coupled to a narrowband high-finesse optical cavity. The trap creates a static harmonic potential $V(x,y,z) = \frac{1}{2}\,m_\mathrm{Rb} \, \omega^2_{x,y,z} \times (x^2,y^2,z^2)$ with $m_\mathrm{Rb}$ being the mass of the rubidium-87 atoms and $a_\mathrm{ho} = \sqrt{\hbar / (m \omega_\mathrm{ho})}$ the harmonic oscillator length. length. In this equation, $\omega_\mathrm{ho} = (\omega_\mathrm{x}\omega_\mathrm{y}\omega_\mathrm{z})^{1/2}$. $\omega_{x,y,z} = 2\pi \times (119.0,102.7,24.7)$~Hz are the measured trap frequencies. The corresponding Thomas-Fermi radii of the ensemble are $(r_x,r_y,r_z) = (3.7, 4.3, 18.1)~\mu m$. These radii are significantly smaller than the size of the Gaussian-shaped pump beam, which has a waist of $w_\mathrm{pump}~\approx~125~\mu m$. The pump beam is oriented transversally with respect to the cavity axis and retro-reflected to form a standing wave potential. The cavity field has a decay rate of $\kappa \approx ~2\pi \times3.2\,$kHz, which is comparable to the recoil frequency $\omega_\mathrm{rec} = E_\mathrm{rec}/\hbar  =~2\pi \times3.7\,$kHz for a pump wavelength of $\lambda_\mathrm{p} = 791.59\,$nm. The pump laser is blue detuned with respect to the relevant atomic transition of $^{87}$Rb at 794.98~nm. The maximum light shift per photon is $U_0 = 2\pi \times 0.7~\mathrm{Hz}$.
\section{Cavity field detection}
Our experimental system is equipped with two detection setups for the light leaking out of the cavity. On one side of the cavity, we use a single photon counting module (SPCM), which provides access to the intensity of the intra-cavity field and the associated photon statistics. On the other side of the cavity, a balanced heterodyne detection setup is installed, which uses the pump beam as a local reference. The beating signal of the local oscillator with the light leaking out of the cavity allows for the observation of the time evolution of the intra-cavity photon number $N_\mathrm{P}(t)$ and the phase difference between the pump and the cavity field.
\section{Identifying the optimal pump parameters}
\begin{figure}[htbp]
	\centering
	\includegraphics[width=0.9\columnwidth]{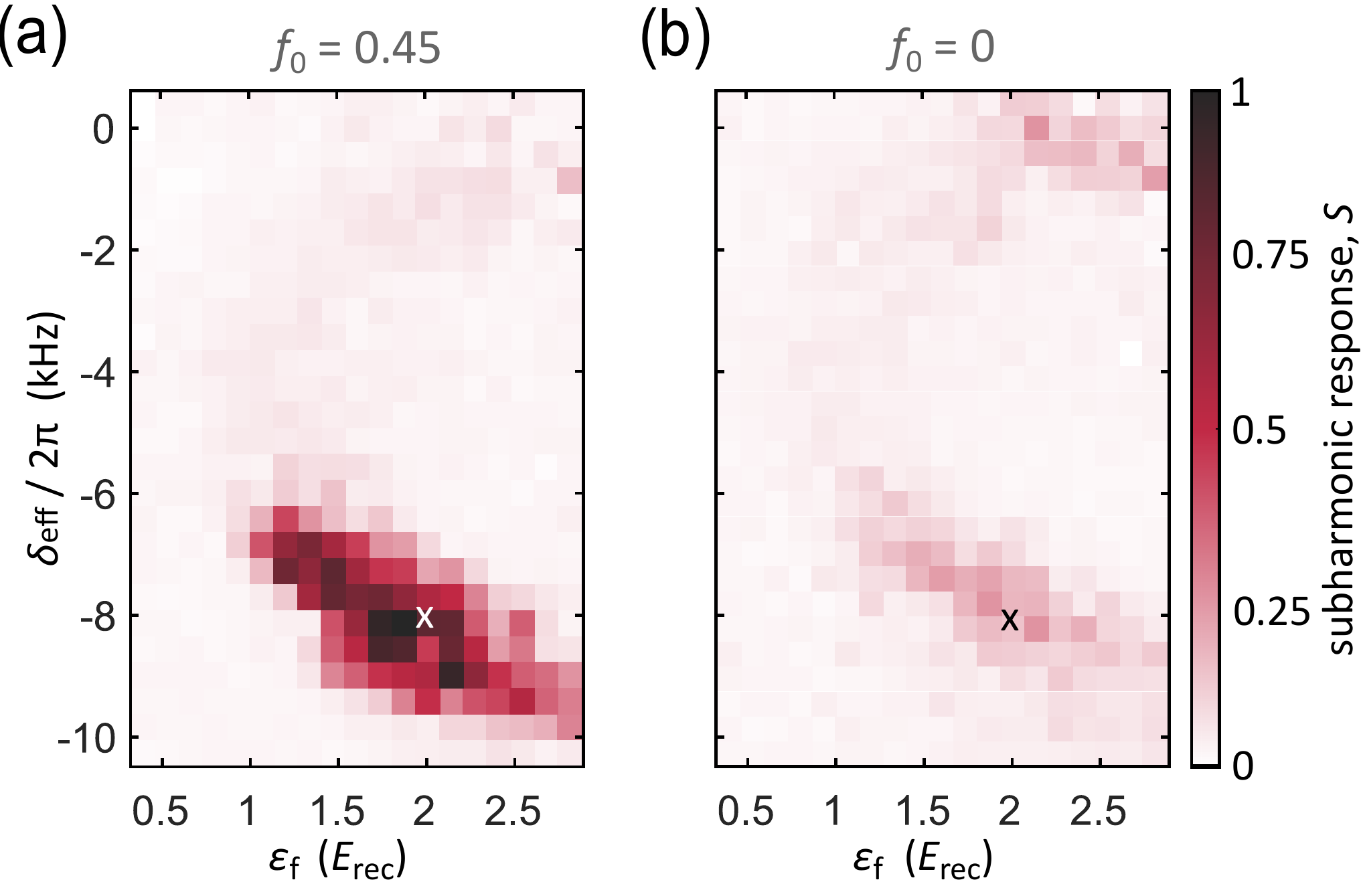}
\caption{Subharmonic response $S$ for the modulated (a) and for the non-modulated case (b), plotted versus the effective pump-cavity detuning $\delta_\mathrm{eff}$ and the mean final pump strength $\varepsilon_\mathrm{f}$. The modulation parameters $\omega_\mathrm{dr}=2\pi\times22.5$~kHz and $f_\mathrm{0}=0.45$ are kept constant for the entire measurement. To obtain (a) and (b), we ramp the pump strength $\varepsilon$ to its final value $\varepsilon_\mathrm{f}$, for a fixed $\delta_\mathrm{eff}$. After a hold time of $0.5$~ms at $\varepsilon_\mathrm{f}$, $f_\mathrm{0}$ is ramped to 0.45 for fixed $\omega_\mathrm{dr}=2\pi\times22.5\,$kHz within $0.5$~ms and subsequently held constant for 10~ms. The evolution of $N_\mathrm{P}(t)$ is recorded during this time interval and its Fourier spectrum is calculated by a discrete Fast-Fourier-Transformation (FFT) method. The amplitude of the Fourier spectrum at $\omega=0.5\,\omega_\mathrm{dr}$, normalized to the maximally observed value of $S$, is plotted according to the shown colour code. The parameter space is divided into 21 $\times$ 27 plaquettes and averaged over 5 to 10 experimental realizations. The white and black crosses indicate the parameter values $\delta_\mathrm{eff} = -2\pi\times8.2\,$kHz and $\varepsilon_\mathrm{f}=2.0\,E_\mathrm{rec}$, which are used for the measurements presented in Figs.~1-4 in the main text of this manuscript.}
\label{sfig:1} 
\end{figure}

The standing wave pump field is characterized by two parameters: the effective pump-cavity detuning $\delta_\mathrm{eff}$ and the time-dependent pump strength $\varepsilon(t)$. The latter follows the equation $\varepsilon(t)=\varepsilon_\mathrm{f} [1 + f_0\cos(\omega_\mathrm{dr}t)]$, with the mean pump strength $\varepsilon_\mathrm{f}$ after ramping is completed, the driving strength $f_0$ and the driving frequency $\omega_\mathrm{dr}$. Fig.~2(a) in the main text shows the dependence of the subharmonic response $S$, used to quantify the IL process, for variable modulation parameters $f_0$ and $\omega_{dr}$. $S$ is the amplitude of the single-sided spectrum at half of the driving frequency, normalized to the observed maximal value. In this section, we hold the driving strength $f_0=0.45$ and the driving frequency $\omega_\mathrm{dr}=2\pi\times22.5\,$kHz constant and identify the parameter regime in the space spanned by $\delta_\mathrm{eff}$ and $\varepsilon_\mathrm{f}$ where IL works most efficiently, and hence, $S$ is maximized. In Fig.~\ref{sfig:1}(a), we observe an elliptically shaped island with strong enhancement of $S$ for large negative detunings, compared to the non-modulated case shown in Fig.~\ref{sfig:1}(b). The value of $S$ increases by almost a factor of five for the optimal parameter set. Furthermore, the modulation leads to the suppression of oscillations at small negative $\delta_{\mathrm{eff}}$. The white and black crosses indicate an optimized set of parameters, i.e., $\delta_\mathrm{eff}=-2\pi\times8.2\,$kHz and $\varepsilon_\mathrm{f} = 2.0\,E_\mathrm{rec}$, used for the measurements presented in Figs.~1-4 in the main text of this manuscript.

\begin{figure}[htbp]
\centering
\includegraphics[width=0.9\columnwidth]{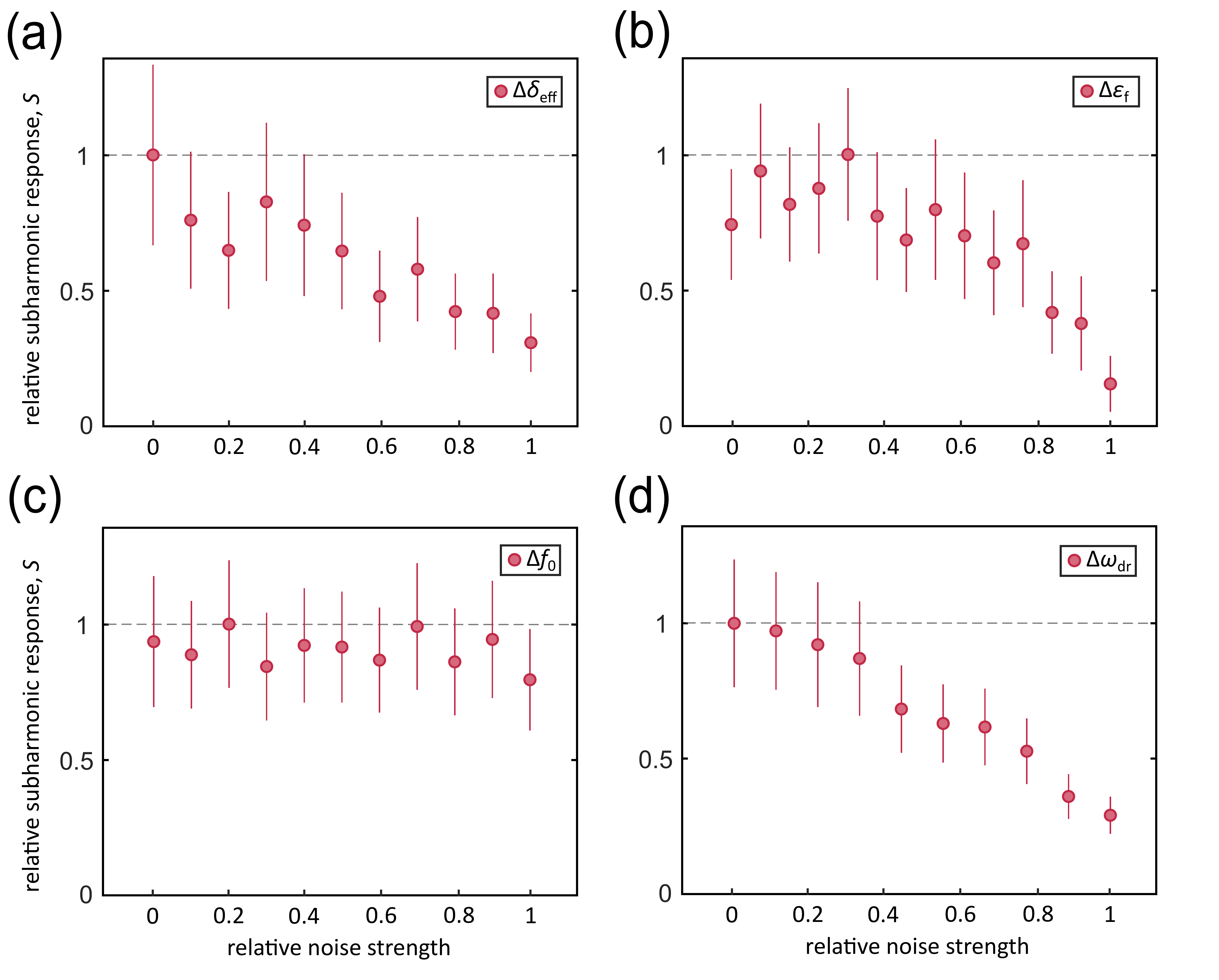}
\caption{(a) Temporal perturbation of the effective pump-cavity detuning $\delta_\mathrm{eff}$. The maximum relative noise strength corresponds to adding white noise with an amplitude of $\Delta\delta_\mathrm{eff,max}=\pm2.5\,$kHz. (b) Temporal perturbation of the final pump strength $\varepsilon_\mathrm{f}$. The maximum relative noise strength corresponds to adding white noise with an amplitude of $\Delta\varepsilon_\mathrm{f,max}=\pm0.5\,E_\mathrm{rec}$. (c) Temporal perturbation of the driving strength $f_\mathrm{0}$. The maximum relative noise strength corresponds to adding white noise with an amplitude equal to $f_\mathrm{0}$. (d) Temporal perturbation of the driving frequency $\omega_\mathrm{dr}$. The maximal relative noise strength corresponds to adding white noise with a deviation of $\Delta\omega_\mathrm{dr}=2\pi\times20\,$kHz. For all measurements, the bandwidth of the white noise was $50\,$kHz, $\delta_\mathrm{eff}=-2\pi~\times8.2$~kHz, $\varepsilon_\mathrm{f}=2.0~E_\mathrm{rec}$, $f_0=0.45$, and $\omega_\mathrm{dr}=2\pi\times22.5\,$kHz.}
\label{sfig:2} 
\end{figure} 

\section{Robustness against temporal perturbations}
In this section, we investigate the robustness of the IL process against temporal perturbations applied to all the parameters that characterize the pump field of the periodically driven atom-cavity system, which are $\delta_\mathrm{eff}$, $\varepsilon_\mathrm{f}$, and the modulation parameters $f_0$ and $\omega_\mathrm{dr}$. For these experiments, we initialize our system in the discrete time crystal (DTC) phase using the following protocol: first, we linearly increase the pump strength $\varepsilon(t)$ to its final value $\varepsilon_\mathrm{f}=2.0~E_\mathrm{rec}$ for fixed $\delta_\mathrm{eff}=-2\pi~\times8.2$~kHz to prepare our system in the continuous time crystal (CTC) regime. After a waiting time of 0.5~ms, followed by a 0.5~ms ramp of the driving strength to $f_0=0.45$ for $\omega_\mathrm{dr} = 2\pi\times22.5\,$kHz, we keep all pump parameters constant for 10~ms and separately add white noise with a bandwidth of 50 kHz to each of them. The subharmonic response $S$ for increasing noise strength is plotted in Fig.~\ref{sfig:2}. We observe robustness of the oscillations for nonzero noise strength for all four pump parameters.   

\subsection{IL for fractions of $\omega_\mathrm{dr}/\omega_\mathrm{CTC}$ close to $1$ and $1/2$}
\begin{figure}[htbp]
\centering
\includegraphics[width=0.9\columnwidth]{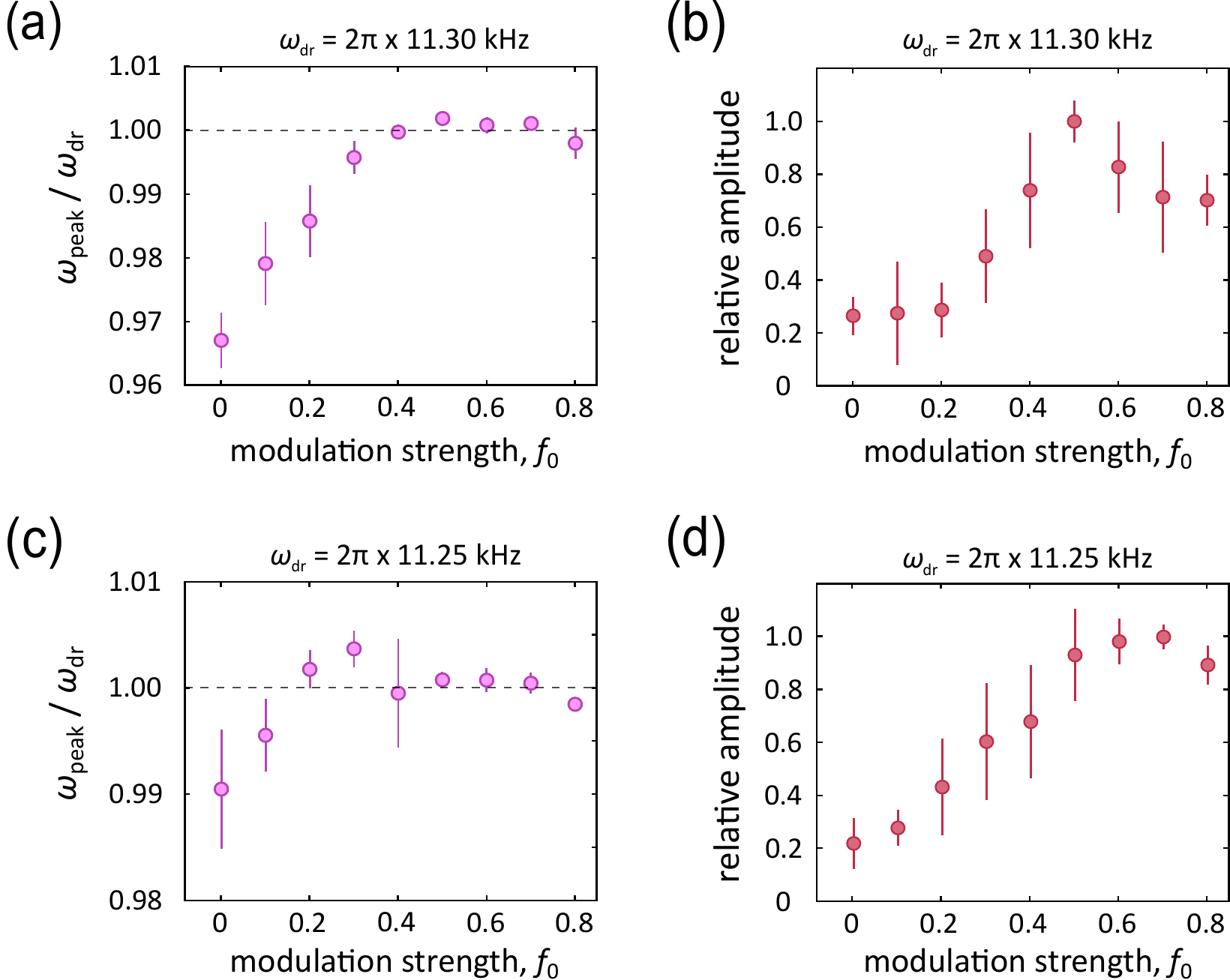}
\caption{(a,~c) Relative response frequencies $\omega_\mathrm{peak}$ for driving frequencies $\omega_\mathrm{dr} = 11.30$~kHz and $\omega_\mathrm{dr} = 11.25$~kHz, respectively. The response frequency $\omega_\mathrm{peak}$ is obtained as the frequency of the dominant spectral component in a Gaussian fit of the Fourier spectrum $N_\mathrm{P}(\omega)$. (b,~d) Relative amplitude of the dominant spectral component at frequency $\omega_\mathrm{peak}$, corresponding to (a) and (c), respectively. Note that the response frequency locks to the modulation frequency. Hence, there is no spontaneous symmetry breaking and no time crystal.}
\label{sfig:3} 
\end{figure}

\begin{figure}[!htbp]
\centering
\includegraphics[width=0.9\columnwidth]{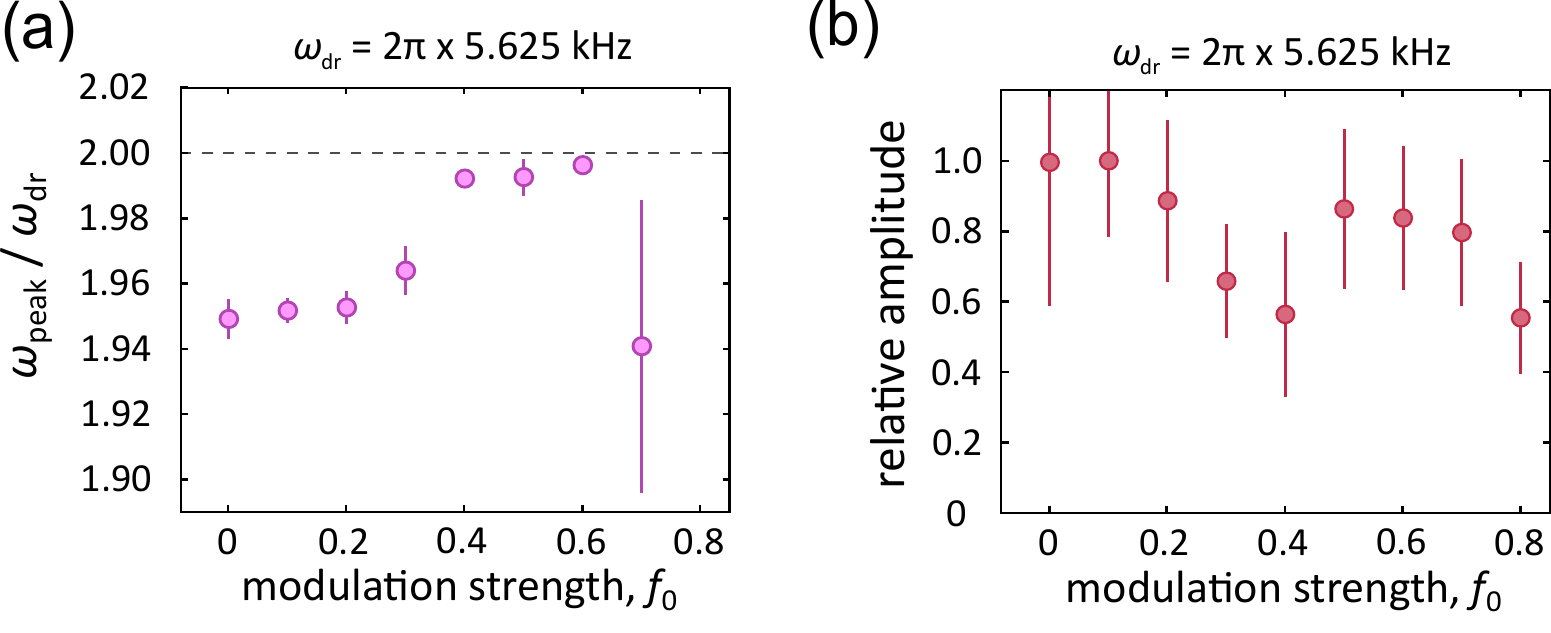}
\caption{(a) Relative response frequency $\omega_\mathrm{peak}$ for the driving frequency $\omega_\mathrm{dr} = 5.625$~kHz, plotted against the driving strength $f_0$. $\omega_\mathrm{peak}$ is obtained as the frequency of the dominant spectral component in a Gaussian fit of the Fourier spectrum $N_\mathrm{P}(\omega)$. (b) Corresponding relative amplitude. Note that the response frequency locks to twice the modulation frequency. Hence, there is no spontaneous symmetry breaking and no time crystal but instead a frequency doubling scenario.}
\label{sfig:4} 
\end{figure}

Next, we investigate the IL process when the ratio between the driving and the intrinsic limit cycle frequencies is close to $1$ or $1/2$, in contrast to the experiments presented in the main text, where this ratio is close to 2. We prepare the system in the CTC regime, such that its intrinsic frequency is around $\omega_\mathrm{CTC} = 2\pi \times 11.10$~kHz. Fig.~\ref{sfig:3} shows the case of $\frac{\omega_\mathrm{dr}}{\omega_\mathrm{CTC}} \approx 1$. The experimental protocol and evaluation method are the same as in Figs.~2(b,c) of the main text. However, for the data shown in Fig.~\ref{sfig:3}, we drive at $\omega_\mathrm{dr} = 2\pi \times 11.30$~kHz (a,b) and $\omega_\mathrm{dr} = 2\pi \times 11.25$~kHz (c,d) to investigate how the CTC is entrained to the driving frequency $\omega_\mathrm{dr}$. See results in Figs.~\ref{sfig:3}(a,b) and Figs.~\ref{sfig:3}(c,d), respectively. \\ Fig.~\ref{sfig:4} shows the IL process when the ratio between the driving and the intrinsic CTC frequencies are close to $1/2$. We drive the CTC at $\omega_\mathrm{dr} = 2\pi \times 5.625$~kHz and observe the response frequency $\omega_\mathrm{DTC}$ entrained to twice the driving frequency. The observations in this section emphasize a key feature of nonlinear systems, i.e., that their limit cycle frequencies can be entrained to assume any rational fraction of the driving frequency. 

\section{Theoretical model}

To theoretically model the experimental results, we only include the degrees of freedom along the pump and cavity axes. In doing so, we consider a 2D system and neglect the short-range contact interaction between the atoms. Thereby, the atom-cavity Hamiltonian in second quantized form comprises three contributions, one from the cavity photons, the atoms, and the light matter interactions
\begin{equation}
\hat{H} =\hat{H}_\mathrm{c}+\hat{H}_\mathrm{a}+\hat{H}_\mathrm{ac}.
\end{equation}
The Hamiltonian for the single-mode cavity is $\hat{H}_\mathrm{c}=-\hbar \delta_{c}~\hat{a}^\dagger\hat{a}$, where $\hat{a}~(\hat{a}^\dagger)$ is the bosonic annihilation (creation) operator for the cavity photons and  $\delta_\mathrm{c}<0$ is the detuning between the pump and cavity frequencies. The atomic Hamiltonian is given by
\begin{equation}
\hat{H}_\mathrm{a} = \int dy dz~ \hat{\Psi}^\dagger(y,z) \left(-\frac{\hbar^2}{2m_\mathrm{Rb}} \nabla^2+\mathrm{V}_\mathrm{ext}(y,z)\right) \hat{\Psi}(y,z),
\end{equation}
where $m_\mathrm{Rb}$ is the mass of a $^{87}$Rb atom. $\mathrm{V}_\mathrm{ext}(y,z)=\varepsilon(t)\cos^2(ky)$ describes the potential due to the standing wave pump beam. The depth of this potential is characterized by $\varepsilon(t)~=~\varepsilon_\mathrm{f}[1+f_0 \sin(\omega_{dr}t)]$ and its wavelength $\lambda_\mathrm{p}$ is incorporated in the wave vector $k=2\pi/\lambda_\mathrm{p}$. The bosonic field operators for the atoms are $\hat{\Psi}(y,z)$ and  $\hat{\Psi}^\dagger(y,z)$. Finally, the light-matter interaction Hamiltonian is given by
\begin{multline}
\hat{H}_\mathrm{ac}= \int dy dz~ \hat{\Psi}^\dagger(y,z) \\
 \left(\hbar U_0\cos^2(kz)\hat{a}^\dagger\hat{a}+\hbar \sqrt{\hbar\varepsilon(t)U_0}\cos(ky)\cos(kz)\left[\hat{a}^\dagger+\hat{a} \right] \right)\\
 \hat{\Psi}(y,z),
\end{multline}
where $U_0>0$ is the light shift per intra-cavity photon. We assume that the wave number of the cavity field is equal to the wave number $k$ of the pump field. 

We simulate the dynamics of the system using a the truncated Wigner approximation (TWA) for open systems \cite{Polkovnikov2010, Cosme2019}. To this end, we first expand the atomic field operators in the basis of plane waves
\begin{equation}
\hat{\Psi}(y,z) = \sum_{n,m} \hat{\phi}^\dagger_{n,m} e^{inky}e^{imkz},
\end{equation}
where the bosonic creation and annihilation operators are $\hat{\phi}^\dagger_{n,m}$ and $\hat{\phi}_{n,m}$, respectively. Within the TWA, the operators are treated as $c$ numbers $\hat{a} \to a$ and $\hat{\phi}_{n,m} \to {\phi}_{n,m} $. The semiclassical equations of motion are
\begin{align}\label{eq:eom}
i\frac{\partial \phi_{n,m}}{\partial t} &= \frac{1}{\hbar}\frac{\partial H}{\partial \phi^*_{n,m}}, \\ \nonumber
i\frac{\partial a}{\partial t} &= \frac{1}{\hbar}\frac{\partial H}{\partial a^*_{n,m}} - i\kappa a + i \xi,
\end{align}
where the fluctuation strength $\xi$ associated to the cavity field decay follows $\langle \xi^*(t)\xi(t')\rangle = \kappa \,\delta(t-t')$. We initialize both the atomic and photonic modes as coherent states and use the appropriate Wigner distribution to sample the initial state for the time evolution according to Eq.~\eqref{eq:eom}. In doing so, we effectively include the leading order quantum corrections to the mean-field predictions, which correspond to a single trajectory and the absence of stochastic noise $\xi=0$. We consider $10^2$ trajectories in our TWA simulations. Furthermore,  we use $\delta_\mathrm{eff} = -2\pi \times 7~\mathrm{kHz}$, $\varepsilon_\mathrm{f} = 1.7~E_\mathrm{rec}$, and $\omega_\mathrm{dr}=2\pi \times 20.5~\mathrm{kHz}$. The remaining parameters are the same as those in the experiment. 

\section{Finite size effects}
To analyze the finite-size behaviour of the system, we vary the particle number $N_\mathrm{a}$ for fixed $N_\mathrm{a} U_0 = 2\pi \times 28~\mathrm{kHz}$. We compare the results of exemplary trajectories within TWA and mean-field theory. The all-to-all couplings of the atoms due to the cavity photons suggest that mean-field theory captures the thermodynamic limit $N_\mathrm{a} \to \infty$, and thus mean-field results provide an idealized scenario for the system. To quantify the stability of the time crystals, we obtain the power spectrum of the intra-cavity photon number $N_\mathrm{P}(\omega)$. We then calculate the relative crystalline fraction defined as the ratio between the maximum peak of the power spectrum of the TWA and mean-field results, $\Xi = \mathrm{max}[N_\mathrm{P,TWA}(\omega_\mathrm{DTC})]/ \mathrm{max}[N_\mathrm{P,MF}(\omega_\mathrm{DTC})]$. This quantifies the stability of the time crystals for finite $N_\mathrm{a}$ relative to the idealized mean-field limit, which for the parameters chosen here exhibits stable oscillations at a well-defined frequency.

\begin{figure}[htbp]
\centering
\includegraphics[width=0.7\columnwidth]{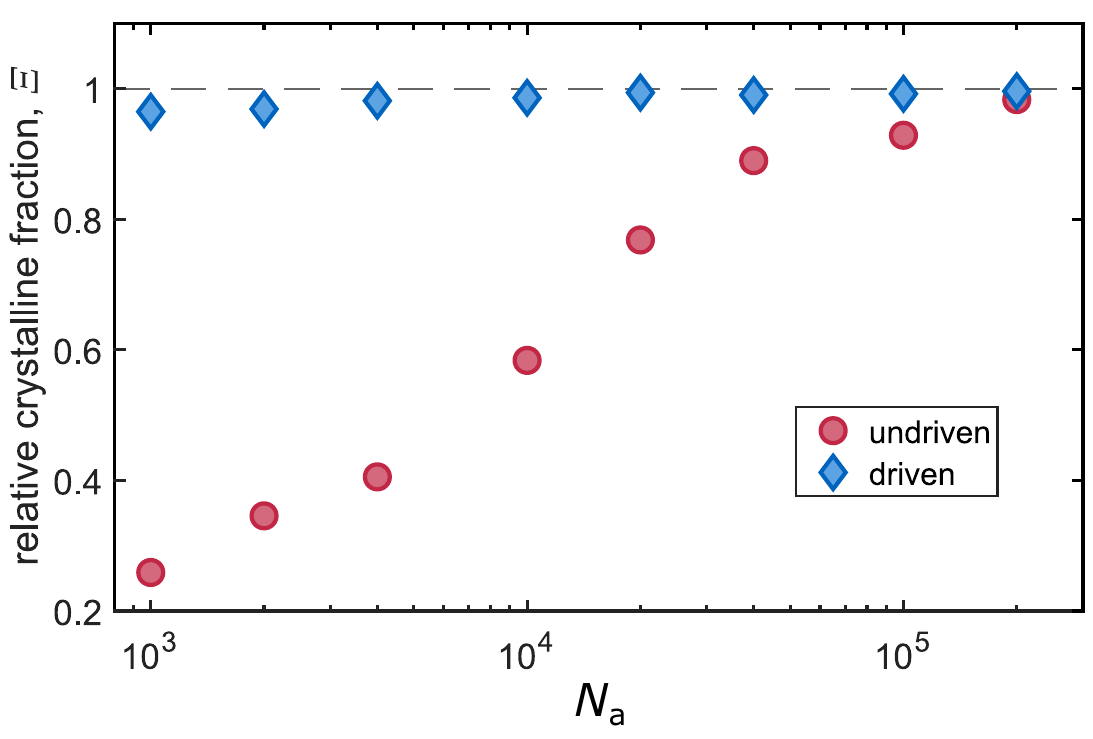}
\caption{Relative crystalline fraction $\Xi$ as a function of the particle number $N_\mathrm{a}$ for the undriven and driven systems. The parameters are $\delta_\mathrm{eff} = -2\pi \times 7~\mathrm{kHz}$ and $\varepsilon_\mathrm{f} = 1.7~E_\mathrm{rec}$. In the driven case, the driving parameters are $\omega_\mathrm{dr}=2\pi \times 20.5~\mathrm{kHz}$ and $f_0=0.15$. All other parameters are identical with those used in the experiment.}
\label{sfig:5} 
\end{figure}

We present in Fig.~\ref{sfig:5} the relative crystalline fraction $\Xi$ for different particle numbers $N_\mathrm{a}$. For both driven and undriven cases, the relative crystalline fraction increases $\Xi$ with $N_\mathrm{a}$ as it approaches the mean-field prediction in the thermodynamic limit $N_\mathrm{a} \to \infty$. This suggests that the oscillation amplitude of the time crystals becomes more stable with increasing $N_\mathrm{a}$.  For small $N_\mathrm{a}$, in which quantum fluctuations become important, the typical values of $\Xi$ for the undriven system are much less than those for the driven system. This further highlights the  capability of periodic driving to enhance the stability of a time crystal. Therefore, in general, the entrained time crystals are more stable than their undriven counterparts.

\section{Atom number fluctuations}
In the experiment, additional fluctuations from technical noise are present when preparing the initial BEC. To study their consequence on the IL process with the periodic drive, we include artificial noise in the initial particle number in our TWA simulations. Specifically, we increase the fluctuations in the occupation of the lowest momentum mode, the BEC mode, by increasing the standard deviation of the Gaussian distribution used for initial state sampling. In the absence of technical noise, the inherent number fluctuations of a coherent state correspond to a standard deviation of $\sigma_{N_\mathrm{a}} = \sqrt{N_\mathrm{a}}$. We model the experimentally observed particle number fluctuation by increasing the standard deviation to a value consistent with the experiment, which is $\sigma_{N_\mathrm{a}} = 10\sqrt{N_\mathrm{a}} = 2\times 10^3$. 

\begin{figure}[!htbp]
\centering
\includegraphics[width=0.9\columnwidth]{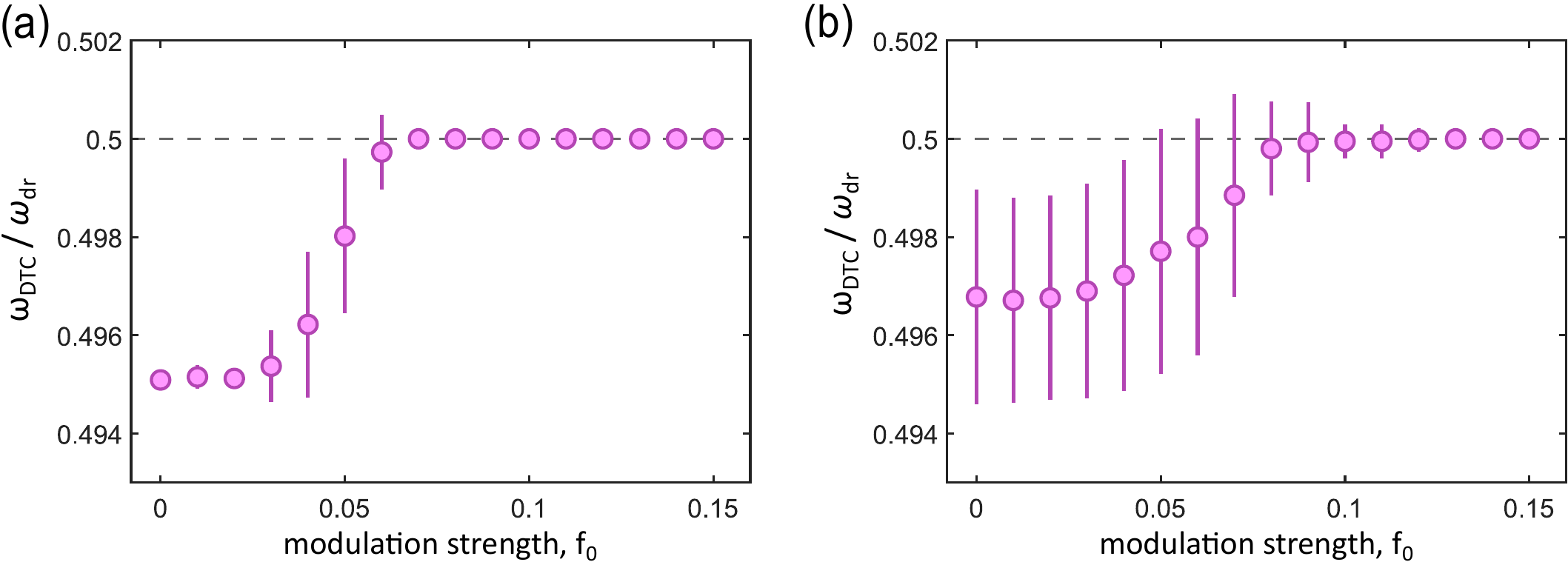}
\caption{Numerical results on the influence of atom number fluctuations on the response frequency. The dominant response frequency $\omega_\mathrm{DTC}$ in units of $\omega_\mathrm{dr}$ is plotted versus the driving strength $f_0$ for a system with (a) only inherent quantum fluctuations and (b) with both quantum and technical noise with a mean atom number $N_\mathrm{a} = 40 \times 10^3$ and $\sigma_{N_\mathrm{a}} = 10\sqrt{N_\mathrm{a} } $.}
\label{sfig:6} 
\end{figure}

In Fig.~\ref{sfig:6}, we show the numerical results comparing the dominant response frequency $\omega_\mathrm{DTC}$ and its shot-to-shot fluctuations, depicted as error bars, for the ideal case with only the inherent quantum fluctuations of the initial BEC (A) and the case with additional particle number fluctuations due to technical noise (B). The shot-to-shot fluctuations of $\omega_\mathrm{DTC}$ are found to be generally larger when there is additional noise, especially for weak driving. However, we find that IL still works not only with regard to locking the signal to a subharmonic of the drive but also in suppressing the associated shot-to-shot fluctuations, albeit for larger driving strength when compared to the ideal scenario, which only includes the inherent quantum noise.

\section{Time evolution of momentum distribution}

\begin{figure}[htbp]
\centering
\includegraphics[width=0.7\columnwidth]{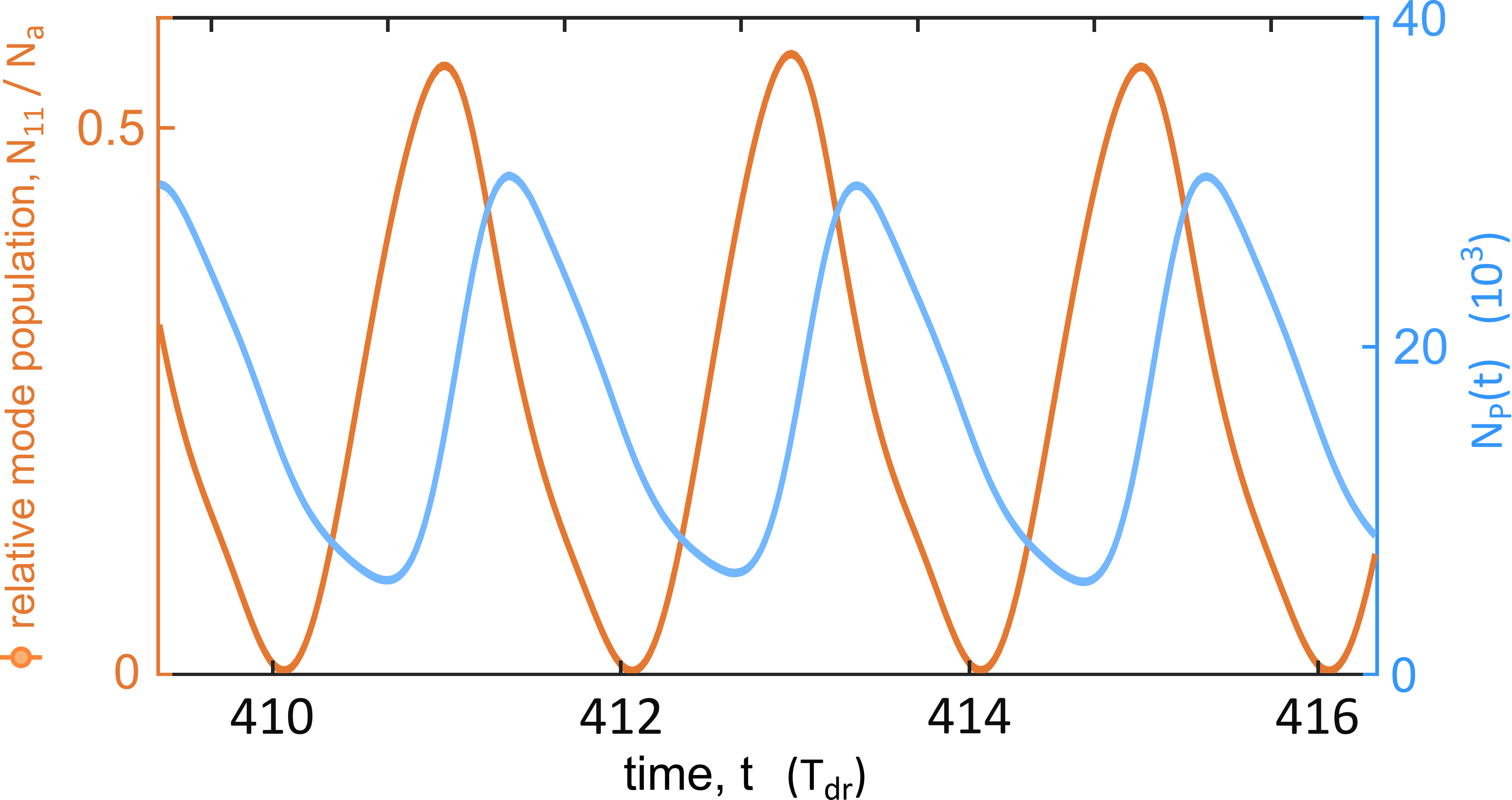}
\caption{Numerical simulation of the time evolution of higher order momentum components. The blue solid line plots $N_\mathrm{P}(t)$ versus time in units of the oscillation period $T_\mathrm{dr}$ of the external drive. The orange solid line shows the corresponding time evolution of $N_\mathrm{11}/N_\mathrm{a}$, where $N_\mathrm{11}$ denotes the sum of the populations of the momentum components $\{p_\mathrm{y},p_\mathrm{z}\} = \{\pm 1,\pm 1\}\hbar k$. The mean atom number is $N_\mathrm{a} = 40 \times 10^3$.}
\label{sfig:7} 
\end{figure}

In Fig.~\ref{sfig:7}, the time evolution of the sum $N_\mathrm{11}$ of populations of the four momentum modes $\{p_\mathrm{y},p_\mathrm{z}\} = \{\pm 1,\pm 1\} \hbar k$, normalized to the total atom number $N_\mathrm{a}$, is shown by the orange line graph. In addition, the time evolution of the intra-cavity photon number $N_\mathrm{P}(t)$ is plotted as a solid blue line. In agreement with the experimental findings in Fig.~4(a) of the main text, a delay between both graphs is found, as expected in the recoil resolved regime present in our atom-cavity system. The simulation closely follows the experimental protocol for DTC preparation. That is, TWA trajectories are calculated in the momentum basis and post-selected to belong to the same symmetry-broken state of the emerging DTC. The momentum spectrum and, thus, $N_\mathrm{11}$ is obtained. The simulation neglects contact interaction and atom loss due to technical heating. 

\begin{figure}[htbp]
\centering
\includegraphics[width=0.6\columnwidth]{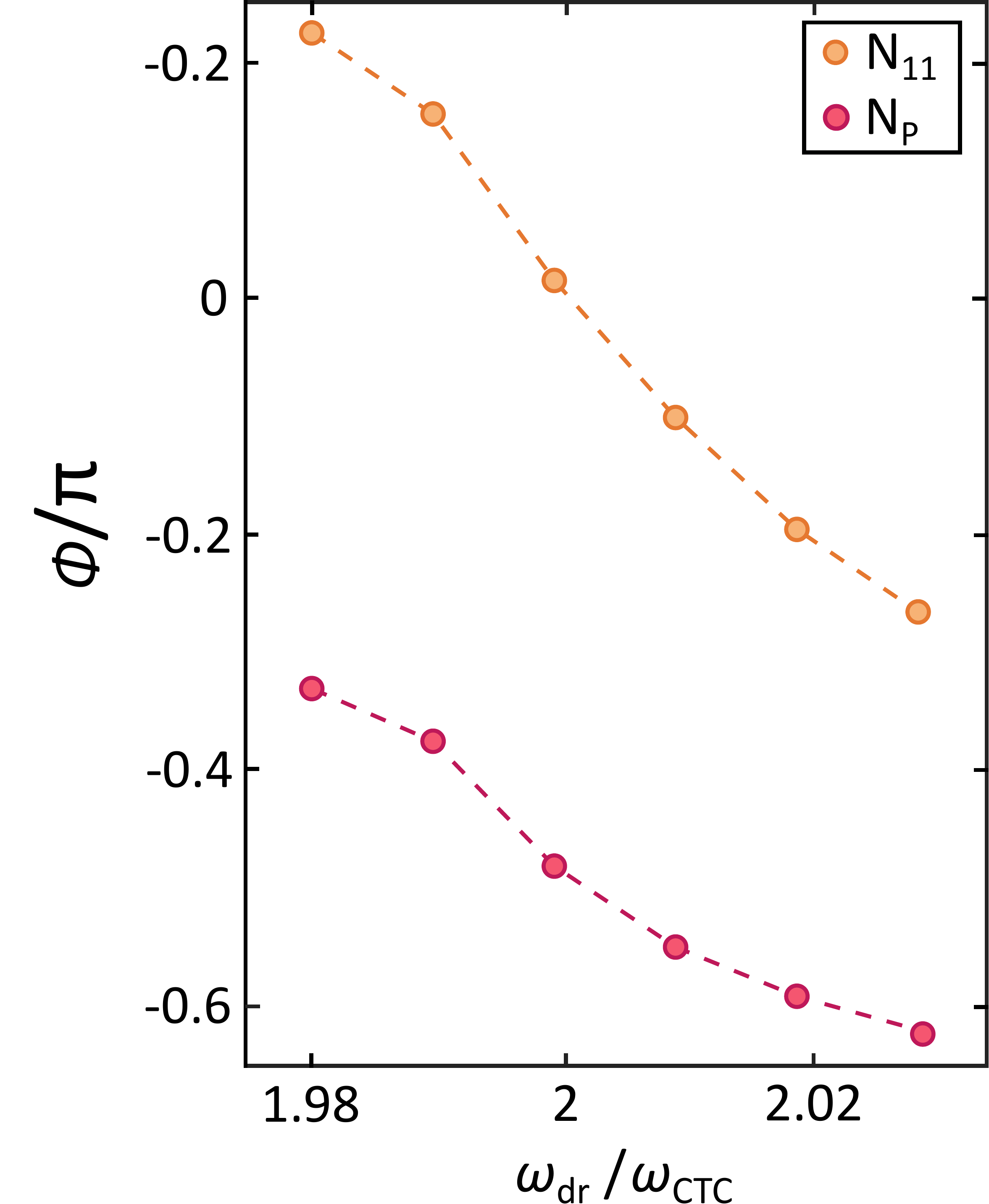}
\caption{Phase delays of matter and light sectors after IL. The calculated phases of the oscillations of $N_\mathrm{P}(t)$ (magenta markers) and $N_\mathrm{11}(t)$ (orange markers) are plotted against the driving frequency $\omega_\mathrm{dr}$.}
\label{sfig:8} 
\end{figure}

In Fig.~\ref{sfig:8}, the calculated phases of the oscillations of $N_\mathrm{P}(t)$ (magenta markers) and $N_\mathrm{11}(t)$ (orange markers) are plotted against the driving frequency $\omega_\mathrm{dr}$ varied across the resonance, where $\omega_\mathrm{dr}$ =  $2 \, \omega_\mathrm{CTC}$. The phase of the drive is set to zero. Note that for $\omega_\mathrm{dr} = 2\,\omega_\mathrm{CTC}$ the time phase of the momentum occupation $N_\mathrm{11}(t)$ becomes the same as that of the drive, which is a characteristic feature of IL. We find similar monotonously decreasing scaling with $\omega_\mathrm{dr}$ for $N_\mathrm{11}(t)$ and $N_\mathrm{P}(t)$. The photon dynamics $N_\mathrm{P}(t)$ show an extra phase lag introduced by the cavity dissipation $\kappa$. These findings are compatible with the observations in Fig.~4(g) of the main text. 

\section{Long-time simulations}
\begin{figure}[htbp]
\centering
\includegraphics[width=1\columnwidth]{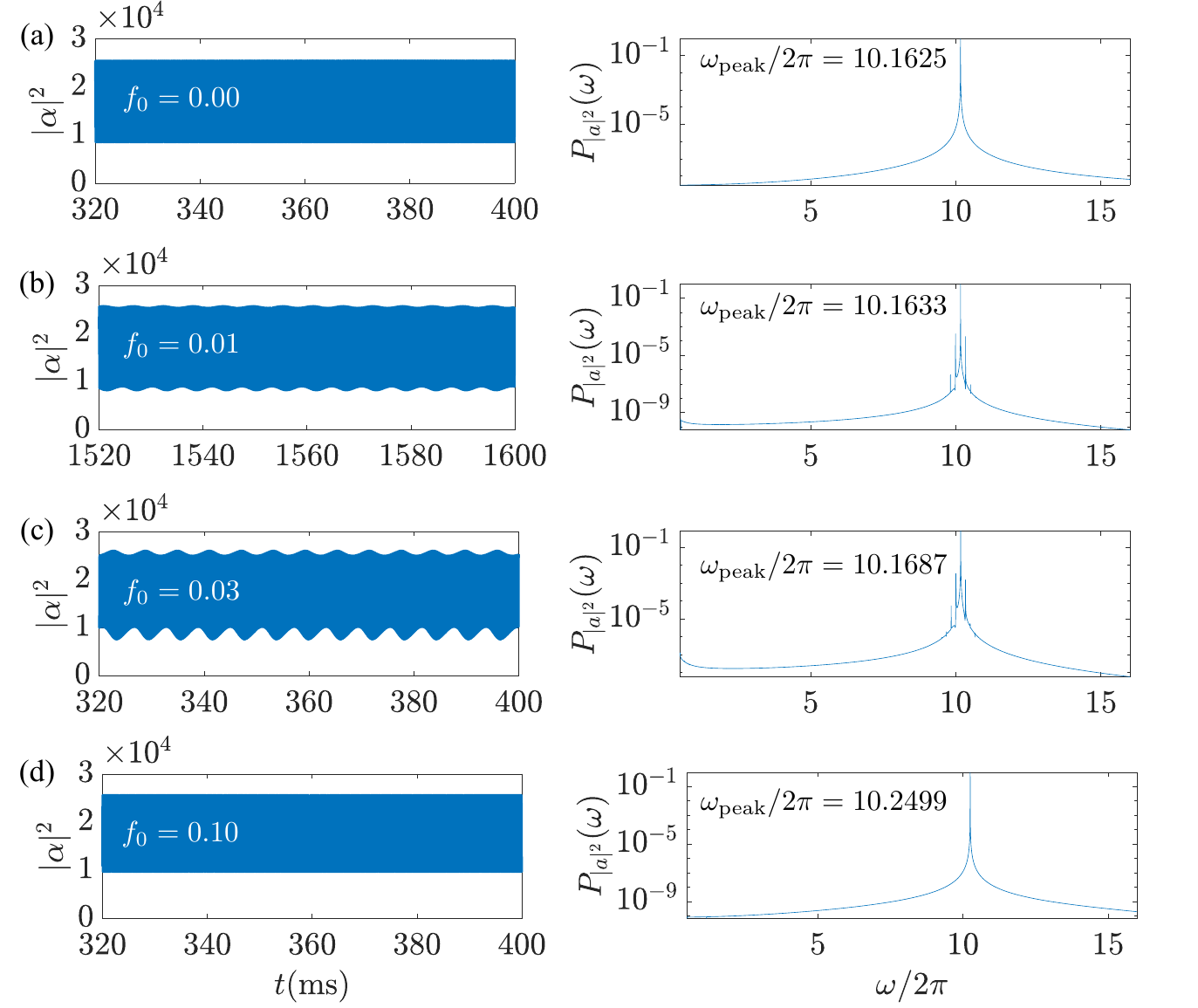}
\caption{Long-time dynamics for different $f_0$ as indicated in each of the plots in the left panel. The right panel shows the corresponding power spectrum with the response frequency also shown in each plot. The remaining parameters are the same as in Fig.~\ref{sfig:5}.}
\label{sfig:11} 
\end{figure}

In Fig.~\ref{sfig:11}, we present some exemplary implementations of the long time dynamics obtained using numerical mean-field simulations. For the CTC case ($f_0=0$) shown in Fig.~\ref{sfig:11}(a), the oscillations have a fixed amplitude, as is indicated by the single peak found in the corresponding power spectrum in the right panel. For weak driving shown in Figs.~\ref{sfig:11}(b) and \ref{sfig:11}(c), quasiperiodic oscillations resulting from the beating between the inherent frequency in the CTC and the driving frequency are present. This can be clearly seen as side peaks in the power spectra, which are incommensurate with the main frequency peak. We note that such quasiperiodic behavior has been considered as time quasicrystals \cite{Aut:18,Giergiel2018}. In the context of nonlinear dynamics and bifurcation theory, the appearance of the quasiperiodic solution signals the formation of an invariant or limit torus \cite{Nicolaou2021}.
The quasiperiodic solution eventually destabilizes for sufficiently large $f_0$, such that the CTC becomes successfully entrained and converted into a DTC represented by $f_0 = 0.10$ in Fig.~\ref{sfig:11}(d). In this case, similar to the CTC, the power spectrum also possesses only a single frequency peak, but now at exactly half the driving frequency. 

While going beyond the scope of this article, it would be interesting to extent the present studies and investigate the influence of thermal and quantum noise on the IL process. For example, it has been found, that noise can induce phase locking and synchronization phenomena in coupled oscillator systems \cite{Braiman1995,Pikovsky1997,Zhou2002,Nicolaou2020}. Similarly, it would be interesting to study the role of quantum noise in the potential stabilization of the limit cycle phase for weak driving and the fate of IL. Another direction of research would be to identify the type of bifurcation that leads to the quasiperiodic behaviour for weak driving using numerical continuation of the limit cycle solution or CTC in the undriven case \cite{Doedel1991,nicolaou_complex_2024}. Finally, the shape of the resonance lobe in Fig. 2 is suggestive of an Arnold tongue, which motivates a more detailed study of a possible mechanism for the period-doubling instability of a limit cycle. To this end, an analysis based on a theory of parametric resonance applied to the DTC in the driven-dissipative Dicke model \cite{Jara2024} could be employed.

\end{document}